# CIM/E Oriented Graph Database Model Architecture and Parallel Network Topology Processing

Zhangxin Zhou[a, b], Chen Yuan[a], Ziyan Yao[a], Jiangpeng Dai[a], Guangyi Liu[a],
Renchang Dai[a], Zhiwei Wang[a], and Garng M. Huang[b]
[a] GEIRI North America, San Jose, CA, USA
[b] Texas A&M University, College Station, TX, USA

*Abstract*—CIM/E is an easy and efficient electric power model exchange standard between different Energy Management System vendors. With the rapid growth of data size and system complexity, the traditional relational database is not the best option to store and process the data. In contrast, the graph database and graph computation show their potential advantages to handle the power system data and perform real-time data analytics and computation. The graph concept fits power grid data naturally because of the fundamental structure similarity. Vertex and edge in the graph database can act as both a parallel storage unit and a computation unit. In this paper, the CIM/E data is modeled into the graph database. Based on this model, the parallel network topology processing algorithm is established and conducted by applying graph computation. The modeling and parallel network topology processing have been demonstrated in the modified IEEE test cases and practical Sichuan power network. The processing efficiency is greatly improved using the proposed method.

## I. Introduction

Energy Management System (EMS) plays a key role in power system operation and control. It includes network topology processing, state estimation, dispatcher power flow, and contingency analysis, etc. The basic functions of EMS typically run in the time frame of minute, which is far behind the sampling rate of SCADA system. To design EMS system in a faster-than real time fashion, the new modeling, algorithms and software platform in parallel are needed.

Nowadays, the relational database is widely implemented in the Energy Management System of power grid control center [1-3]. It is used to store SCADA data, PMU data as well as to integrate with other applications in EMS system.

However, as huge amount of data are acquired and the data size increases, the traditional relational database may not be the most efficient way to handle the real-time analysis and fit into parallel computation applications [4].

The graph database and graph computation have shown their advantages of solving large-scale data management and applying distributed parallel computation algorithm [5]. The vertex and the edge in the graph can act as both a parallel storage unit and a computation unit. The vertices could send computing messages to each other through the communication through edges [6, 7]. Coordinated collaboration and local computation are utilized in a graph computing, which provides an unprecedented solution to do large-scale parallel computing.

To design next-generation Energy Management System software with graph computation technique, the most fundamental step is to model the power grid data into the graph database. In power grid control center, two basic configurations are used to describe the power system models: node breaker model and bus branch model. The monitoring and processing of power system topology in real time are significant for other EMS applications, such as state estimation, contingency analysis, operations and controls [8]. Through network topology processing, a node breaker model is mapped to a bus branch model. .

The CIM/E is an efficient electric grid model exchange standard format developed by State Grid Corporation of China (SGCC). The CIM/E model simplifies CIM/XML model by ignoring terminal without compromising information to process network topology. This paper firstly models the CIM/E standard data into graph database and implements parallel topology processing algorithm using graph computation, which greatly reduces the network topology processing time.

Section II introduces the CIM/E model and compares it with CIM/XML model. Section III and IV discuss the modeling CIM/E into graph database and advantages of graph database for power system EMS applications. The last two sections talk about the parallel topology processing algorithm and case studies.

## II. Introduction of CIM/E Model

### A. Comparison of CIM/E and CIM/XML

Common information model was developed by EPRI for representing power system components and it is used as an open standard for data exchange between Energy Management System vendors. The power system model is exchanged using CIM/XML format. In 2011, SGCC developed model exchange format CIM/E, which has smaller size and simpler form [9]. Due to the smaller size of CIM/E standard data, the efficiency and speed are increased during online data exchange.

For CIM/XML, the topology relationship is described by equipment, terminals and connectivity nodes as shown in Fig. 1.

This work was supported by State Grid Corporation of China technology project SGSHXT00JFJSI700138.

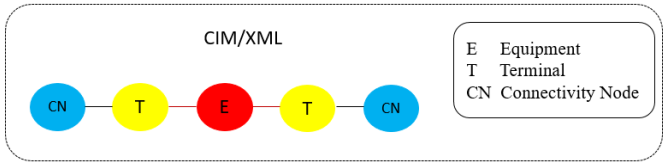

Fig. 1. CIM/XML Model

In CIM/E, the topology relationship only contains equipment and connectivity node as shown in Fig. 2, which can greatly reduce the size as there are huge amounts of terminals in power network. This paper models CIM/E data in graph database, in which the data size is greatly reduced compared with the modeling in S.A. Khaparde's work [10].

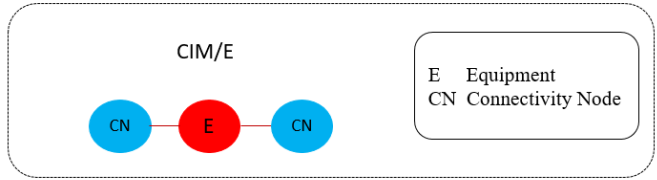

Fig. 2. CIM/E Model

### B. CIM/E Model Structure

CIM/E describes all the objects within one class together in one table, which is efficient and user-friendly. One class table starts with <Class name :: name> and ends with </Class name :: name> . The basic structure of CIM/E language is shown in Fig. 3.

```
<! System Statement !>

<Class Name :: name>
@ Attribute name 1    Attribute name 2    Attribute name 3
// comment
#  Object 1 value 1    Object 1 value 2    Object 1 value 3
#  Object 2 value 1    Object 2 value 2    Object 2 value 3
#  Object 3 value 1    Object 3 value 2    Object 3 value 3
#  ……
</Class Name :: name>

……
……
```

Fig. 3. Basic structure of CIM/E language

## III. CIM/E MODEL IN GRAPH DATABASE

Modeling CIM/E data into graph database is the first and most fundamental step to apply graph database and graph computing to provide solution for power system problems. In this section, we compare two modeling methods. One method is to model objects using vertices only, another is to model objects using vertices and edges. In section IV, the network topology processing will be applied on these two graph-based node breaker models.

### A. All Objects Modeled by Vertices

In the Fig. 4 (a), the one-line diagram of a typically substation model is illustrated, which consists of bus bar, circuit breaker, disconnector, load and generator. The figure 4 (b) shows a CIM/E representation of this substation.

The Fig. 5 shows the modeling of CIM/E standard data in graph database TigerGraph. All the component objects are modeled by vertices in this method. This method could give the operation engineers a straightforward substation visualization and provide them with easy access to conduct the data management and graph model manipulation.

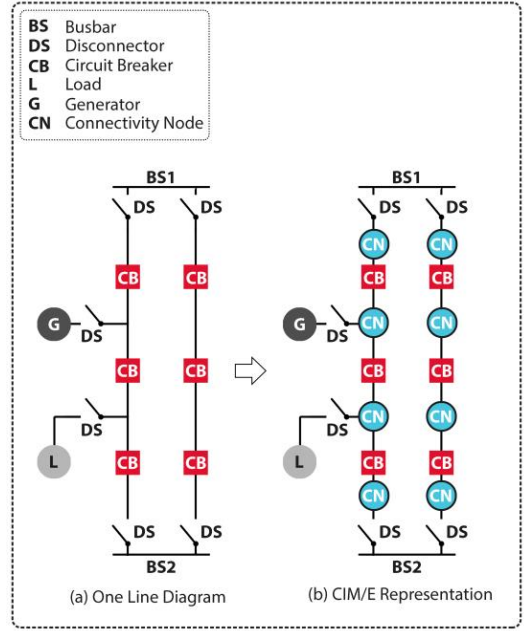

Fig. 4. Substation representation in one-line diagram and CIM/E

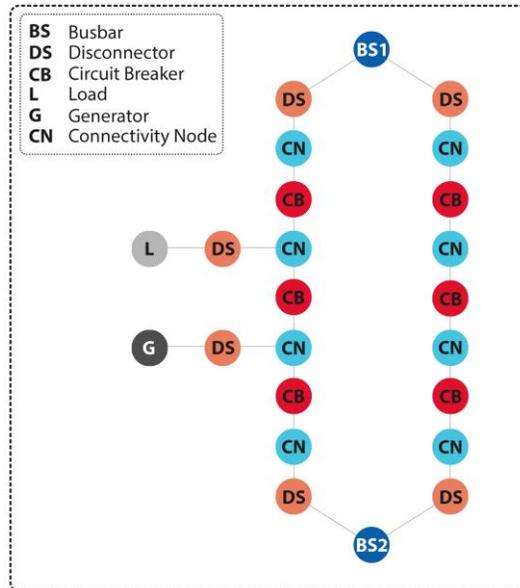

Fig. 5. Substation Modeling in CIMGDB with objects described by vertices

## B. Objects Modeled by Vertices and Edges

In order to reduce the vertex set and prepare for the parallel topology processing, we could also model the circuit breaker and disconnector by edge. All objects are defined using the graph concept G (V, E). The objects that do not have on and off status are defined as vertices. Circuit breaker and disconnector, which have on and off status, are defined as edges. Both the attributes of vertex and edge can be accessed using graph database TigerGraph [7]. For this modeling, the searching steps of topology processing will be tremendously reduced, which will be described in detail in section V.

The Fig. 6 shows this modeling representation in graph database TigerGraph.

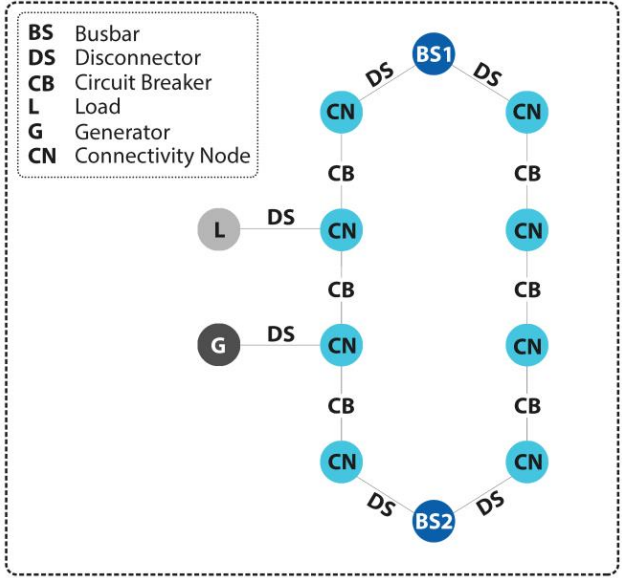

Fig. 6 Substation modeling in CIMGDB with objects described by vertices and edges

## C. Software Package to convert CIM/E into CIMGDB

The basic CIM/E files contain 15 tables, which are base voltage, substation, bus bar, AC line, generator, transformer, load, compensator_P, compensator_S, converter, DC line, island, toponode, breaker, and disconnector tables. A Python package is developed to convert this file to the loading files for graph database. The attribute nd in CIM/E represents the numbering of connectivity nodes. So all the nd columns for connectivity nodes can be found by scanning the CIM/E data file. Then all nds are merged in one csv file and the duplicate information can be deleted. This new file will be the vertex definition of the connectivity node.

Regarding the vertex definition, the key information to be extracted is the connectivity information of an object. For all the conducting equipment that are defined in CIM/E, the neighbors are connectivity nodes. The bus bar, generator, load and compensator_P are special kinds of connectivity nodes with only single side connection.

Each type of object is defined as one type of vertices, which means objects in one type share the same schema for the associated vertex type. We save all those objects information in vertices. For the edge, its category depends on the information of the vertices which are connected through this edge. Once the categories of two vertices are given, the category of the edge is defined. After loading all the files which contain vertices and edges, a query is applied to print the model in visualization window.

## IV. ADVANTAGES OF GRAPH DATABASE FOR POWER SYSTEM APPLICATION

Due to the graph structure and interconnection of the power system, the graph database is well-suited for its modeling, computation, analysis, and visualization. The typical advantages are: 1) efficient data storage for online application; 2) flexible CIM/E modeling in graph database and data management; 3) powerful tool to perform distributed graph computation; 4) seamless connection to the visualization platform.

*1) Efficient data storage in online application:* The data is stored in a graph structure, which is similar to the power grid connection structure. So the query time will be significantly reduced to perform traversal, which is the fundamental operation for topology processing.

*2) Flexible CIM/E modeling in graph database and data management:* The traditional relational database has to do One-to-Many, Many-to-One and Many-to-Many procedure to explore the CIM/E power system data to create system connectivity matrix. Adding and removing object require to rebuild the connectivity matrix. In contrast, the model in graph database can be directly changed when adding and removing nodes and/or edges. The attributes associated to node and edge can be conveniently added and removed.

*3) Powerful tool to conduct distributed graph computation:* In graph database, the vertices are both data storage units and dynamic computation unit[5]. Graph database query supports multiple nodes to access to their neighbors simultaneously by exchanging information through edges and carries out parallel computation using graph computation engine.

*4) Seamless connection to the power grid visualization:* The graph database provides visualization tool to exhibit, power grid attributes and results on the system nature structure. This benefits operators to monitor system online and make decisions intuitively.

## V. PARALLEL NETWORK TOPOLOGY PROCESSING

Network topology processing is usually conducted in two levels, substation level topology processing and then network level topology processing. Substation topology processing is proceeded one substation by one substation by checking the updated status of circuit breakers and disconnectors in the substation.

Actually, the substation topological information is independent from each other. Thus, we could do parallel network topology processing after we model them into graph database using the graph computing engine of TigerGraph, in which the substation level parallel processing is achieved shown in Fig. 7.

The breaker and disconnector statuses are described in CIM/E file and they contain all the connectivity information for power network current status.

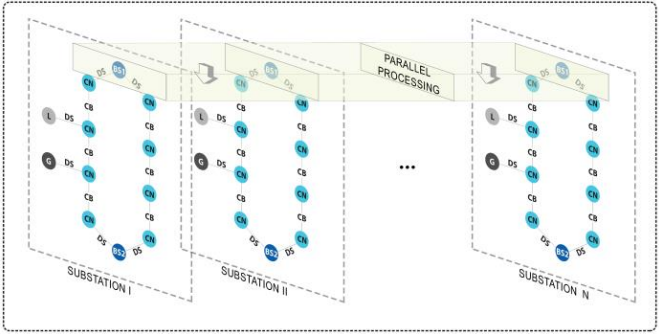

Fig. 7 Parallel Network Topology Processing

*1) Substation Network Topology Processing:* the target of substation network topology processing is to form the topology node which reflect the collection information of objects of substation.

The basic idea of parallel topology processing is to start from the bus bar vertices of substation and to find its neighbor vertices through the connected edges. Because all the bus bars from the substations are the same type of vertex, they will start the search simultaneously. If the status information is shown as connected, then we also assign the bus bar's identification to its neighbor vertex. And we continue to search the vertices that are not visited until all the vertexes of the substation have assigned bus bar identification. The detailed algorithm is presented in following algorithm A.

---

**Algorithm A:** CIM/E oriented Parallel Substation Topology Processing

---

**while** (all bus bar vertices are not processed)
    Start from bus bar vertices of all substations (one bus bar per substation)
  **while** (all vertices are not processed) **do**
    Find neighbor vertices or edges that are not processed
    **if** (vertices or edges are circuit breakers or disconnectors)
      check the connection status
        **if** (the status is connected)
          pass the bus bar id to the vertices or edges
        **end**
    **end**
    **if** (vertices or edges are other type of nodes)
      pass the bus bar id to the vertices or edges
    **end**
  **end**
**end**
**Insert topology nodes according to the IDs generated**

---

The algorithm A could be applied to both modeling methods described in Section III. The performance of the two methods is shown in Section VI.

*2) Network Level Topology Processing:* the network level of topology processing is to describe the connectivity information between different substations. The object AC line in CIM/E have two attributes showing on or off status at the both ends. Starting from the AC lines (and DC if the system has any) in parallel, the neighboring connectivity nodes are searched and topology edges are inserted to connect the topology nodes. The topology nodes and topology edge are buses and branches respectively in the system bus-branch model.

The detailed algorithm is shown in algorithm B.

---

**Algorithm B:** CIM/E oriented Parallel Network Level Processing

---

Start from all the AC line vertices or edges (depends on previous modeling method)
**if** (statuses at both sides are closed)
Find their connectivity nodes IDs generated in **algorithm a** and store them on its attributes
**end**
Insert topology edge simultaneously according to the IDs stored in AC line vertices or edges

---

## VI. CASE STUDY AND RESULTS

### A. NTP on IEEE Test cases

The IEEE 14 and IEEE 118 Bus-Branch model are applied for the case study [11]. The node breaker model are created by using the substation configuration shown in Fig. 4. Then the CIM/E files are created for testing.

To load CIM/E model to graph database, a python package is developed to convert the CIM/E file to the loading files readable by graph database TigerGraph. Using the graph-based node breaker model, the proposed parallel network topology processing algorithm is implemented using graph processing engine.

Two modeling structures in graph database are processed respectively using the methods described in section IV. The network topology processing (NTP) execution time is presented in Table I with all objects modeling by vertex in graph database TigerGraph (the first modeling method).

TABLE I. NTP QUERY RESULTS WITH ALL OBJECTS MODELED BY VERTICES

| IEEE Test Cases | Total Vertices in Graph Database | Total Edges in Graph Database | NTP Execution Time (ms) |
|---|---|---|---|
| 14 | 356 | 376 | 15.2 |
| 118 | 3018 | 3204 | 24.4 |

When the circuit breakers and disconnectors are modeled by edge (the second modeling method), the study results are shown in Table II.

TABLE II. NTP QUERY RESULTS WITH OBJECTS MODELED BY VERTICES AND EDGES

| IEEE Test cases | Total CIM/E Vertices | Total CIM/E Edges | NTP Execution Time (ms) |
|---|---|---|---|
| 14 | 168 | 188 | 10.7 |
| 118 | 1416 | 1602 | 15.9 |

Compared with the results in paper [8], network topology processing time by the methods proposed in this paper using graph processing engine TigerGraph is reduced.

*B. NTP on Sichuan Network*

A practical network (Sichuan, China, which is provincial part of the Chinese power grid) is implemented into graph database TigerGraph to evaluate the parallel topology processing method.

Following the proposed procedure, two modeling methods described in section III are conducted and the NTP execution time results are shown in Table III.

TABLE III. NTP QUERY RESULTS FOR SICHUAN NETWORK

| Sichuan Network | Total Vertices in Graph Database | Total Edges in Graph Database | NTP Execution Time (ms) |
|---|---|---|---|
| Modelling objects using vertices | 78411 | 75277 | 150 |
| Modelling objects using vertices and edges | 48118 | 44984 | 100 |

VII. CONCLUSION

The future power grid control center needs fast simulation tools and algorithms to achieve the real-time data analysis and decision making. This paper introduces the graph database and graph computation to model the power system data and conduct the topology processing. The CIM/E standard data are implemented to demonstrate the modeling in the graph database. The detailed modeling methods based on the graph concept are discussed. The modeling architecture in graph database has the advantages of efficient data storage, flexible data management, and seamless connection to the visualization tools, and is adaptive to the distributed graph computation. The merits mentioned above make the graph database a promising and powerful tool for the next generation Energy Management System.

The parallel network topology processing algorithm based on graph computation is developed and successfully applied in the different test cases. The NTP time is greatly decreased using the proposed method.

ACKNOWLEDGMENT

This work was supported by State Grid Corporation of China technology project SGSHXT00JFJSI700138.